# A Brain-Computer Interface Data Persistence System for Multi-Scenario and Multi-Modal Data: NeuroStore

Yang Chen[1†], Hongxin Zhang[1,2†], Guanyu Xiong[3], Chenxu Li[1], Chengcheng Hong[1], Chen Yang[1*]

1. *School of Electronic Engineering, Beijing University of Posts and Telecommunications*
2. *Beijing Key Laboratory of Work Safety Intelligent Monitoring, Beijing University of Posts and Telecommunications*
3. *School of Information and Communication Engineering, Beijing University of Posts and Telecommunications*

With the rapid advancement of brain-computer interface (BCI) technology, the volume of physiological data generated in related research and applications has grown significantly. Data is a critical resource in BCI research and a key factor in the development of BCI technology, making efficient storage and management of this data increasingly vital. In the realm of research, ample data can facilitate the development of novel algorithms, which can be more accurately validated. In terms of applications, well-organized data can foster the emergence of new business opportunities, thereby maximizing the commercial value of the data. Currently, there are two major challenges in the storage and management of BCI data: providing different classification storage modes for multi-modal data, and adapting to varying application scenarios while improving storage strategies. To address these challenges, this study has developed the NeuroStore BCI data persistence system, which provides a general and easily scalable data model and can effectively handle multiple types of data storage. The system has a flexible distributed framework and can be widely applied to various scenarios. It has been utilized as the core support platform for efficient data storage and management services in the "BCI Controlled Robot Contest in World Robot Contest."

Keywords: brain-computer interface (BCI), data persistence, Brainformatics, Multi-Modal Data

# 1. Introduction

With the continuous advancements in neuroscience, signal processing, and computer technology, brain-computer interface (BCI) technology has experienced significant growth in recent years and has transitioned from pure scientific research to practical applications. EEG-based BCI technology has demonstrated exceptional potential for applications in medical rehabilitation, game development, and the military. Currently, BCI systems are extensively used in rehabilitation medicine for the prevention, intervention, and treatment of diseases such as stroke [1], epilepsy prediction [2], sleep disorders [3], and Alzheimer's disease [4]. Additionally, BCI technologies have been employed to assist patients with motor limitations to interact with the external world [5], such as through the control of wheelchairs [6], robotic arms [7], and spelling [8, 9]. In the realm of gaming, numerous BCI games have been developed based on motor imagery and P300 paradigm [10].

In BCI applications, it is necessary to gather information at various stages of usage and process it through aggregation, computation, and sorting to extract its value. Furthermore, BCI research is becoming increasingly systematic and standardized, and data constitutes a critical resource in research. As a result, the standardized classification, sorting, and label management of long-term accumulated data will become essential. If there were a data platform that offered data management and processing functions, it could leverage the business value of data to drive business growth in terms of application, and in scientific research, it would reduce the difficulty of researchers' work, minimize the probability of errors in data management, and enhance the efficiency of research.

As BCI research continues to advance, the volume of physiological data being generated increases. Many research projects entail the collection of substantial amounts of data, such as the Human Connectome Project (HCP), a global neuroscience community initiative that has collected neurophysiological data from 1200 subjects [11], and the European Union Human Brain Project (HBP), which involves the participation of 117 European institutions [12]. These initiatives have produced normalized datasets that contain large amounts of subject data. The efficient storage and effective management of the large volume of physiological data has become a critical concern in the course of these projects. Since 2019, the Chinese Institute of Electronics has held several editions of the "BCI Controlled Robot Contest in World Robot Contest" [13], which collects data from multiple subjects and evaluates the algorithms of each team online. During the competition, the EEG data of multiple subjects and the detection results of the teams are recorded in real-time into the data persistence system, leading to significant throughput pressure and concurrency requirements. To address these challenges, this study presents NeuroStore, a BCI data persistence system that is designed for multi-scenario and multi-modal data, utilizing a distributed architecture and supporting highly concurrent data storage access. NeuroStore is not only suitable for data development in large projects, but also for data unification management within laboratories or departments and data management in long-term research projects. Further information on NeuroStore can be found at https://github.com/cynotecy/NeuroStore.

Section 2 of this paper outlines the features of NeuroStore, and Section 3 provides a detailed description of its design, including the design of the data model and interfaces. Section 4 conducts comprehensive performance tests to evaluate the system's ability to support multiple concurrent users and store large amounts of data in real-time. Section 5 describes the application of the system in a BCI competition, highlighting its role in data management and analysis and its real-time performance during the competition. Through this paper, the reader will obtain a comprehensive understanding of the design, implementation, and application of NeuroStore, providing a valuable reference for research and practice in similar fields.

## 2. System Overview
### 2.1 Design Objectives
As BCI technology continues to advance, the size and complexity of data generated also increases, presenting a significant challenge in terms of data storage and management. The NeuroStore data persistence system aims to address this challenge by offering a standardized storage approach and a unified data model, facilitating simple and efficient management of large amounts of data and supporting data sharing across different institutions. This system provides researchers with an efficient, reliable, and flexible platform for data management.

Moreover, NeuroStore is designed to meet the demands of both engineering applications and scientific research. It is anticipated to accelerate the transfer of scientific research results into practical applications and contribute to the industrialization of BCI technology, ultimately promoting the advancement of BCI technology.

### 2.2 System Functions
The NeuroStore data persistence system aims to address the challenges posed by the growing volume of data in BCI research by offering the following capabilities:

a) Concurrent Streaming Storage: The system is capable of storing data from one or more data sources without interruption, even in cases where the data sources lack clear end boundaries. It ensures the correctness and consistency of the stored data.

b) Concurrent Query: The system provides the capability for concurrent querying, enabling users to make multiple, simultaneous queries. This capability can handle high concurrency scenarios, such as serving as a data center for multiple processes to meet users' query needs efficiently.

c) Multi-modal Data Management: The system can store data of various modalities, including physiological data, voice, pictures, and other data types. It also stores descriptive information of multi-modal data, such as labels and metadata, allowing users to quickly locate the desired multi-modal data by retrieving and aggregating the descriptive information.

d) Process Information Management: The system can store information related to the BCI usage process, such as subject information and experimental result information, and establish correlations between this information. This enables users to jointly use this information when processing the data to obtain more scientifically meaningful findings.

### 2.3 System Features
a) System Universality:
  1) Model Universality: NeuroStore employs a generic data model, which is based on the BIDS data specification [14] and OHBM's COBIDAS MEEG specification [15]. This standardizes the description of common data in BCI studies, such as EEG (Electroencephalogram) and MEG (Magnetoencephalography), as well as other study information, such as phase results and relevant personnel. The common data model makes NeuroStore suitable for a wide range of BCI projects.
  2) API (Application Programming Interface) Universality: NeuroStore's API is developed in Python, a widely used programming language, and can capture various types of data, including EEG, images, and eye tracking. NeuroStore also offers interfaces with various query functions and can perform complex queries by combining multiple simple interfaces to meet diverse query requirements. NeuroStore can store and query common data in BCI research to support subsequent scientific analysis and can be utilized as a data center by other BCI systems to provide data services for other modules.

b) Flexibility: NeuroStore adopts a component-based design approach and a horizontally scalable architecture, allowing its performance and storage capacity to be expanded by adding new hardware devices. This enables NeuroStore to better match project requirements while reducing the overhead of smaller projects.

c) Ease of Use: NeuroStore is deployed using containerization [16, 17] technology, which enables users to deploy system components with a single click after making a few configuration changes. The system does not require high-performance servers or specialized software knowledge for deployment and management, and provides concise and easy-to-understand interfaces that are accessible to interested researchers.

## 3. System Design
### 3.1 Data Model Design

A data model is a crucial component of system design that regulates the representation, storage, and manipulation of data. The data model proposed in this paper divides all data into five topics: process, data, person, device, and paradigm, as illustrated in Fig. 1. This division is based on the practical needs of BCI research and applications, and facilitates standardized management and efficient utilization of data

a) Process Topic: This topic describes processes such as experimental processes and business processes. Processes are logically important concepts in BCI research, and abstracting process concepts in the data model design reduces the complexity of the model design and improves its practicality and ease of use. The model forms a multi-layered tree structure through self-association to describe process nesting relationships in real-world situations.

b) Data Topic: This topic describes the data stored in the process, including structured data (e.g. result scores) and unstructured data (e.g. EEG). Among the unstructured data is physiological data, which is critical for both research and business applications and is the main focus of data storage and management.

c) Person Topic: This topic describes the persons in the system, including subjects, experimenters, and users. Persons are the primary research objects in scientific research and the main service objects in market applications, and are a vital component of the data model.

d) Device Topic: This topic describes common physiological data acquisition devices, computer devices, and other external controllable devices in the BCI system. BCI technology connects the brain to various devices, and recording device information in BCI research and applications will enhance the overall scenario.

e) Paradigm Topic: This topic aims to accurately and comprehensively describe paradigms in BCI, such as SSVEP [18, 19], ERP [20], and MI [21], to facilitate a deeper study.

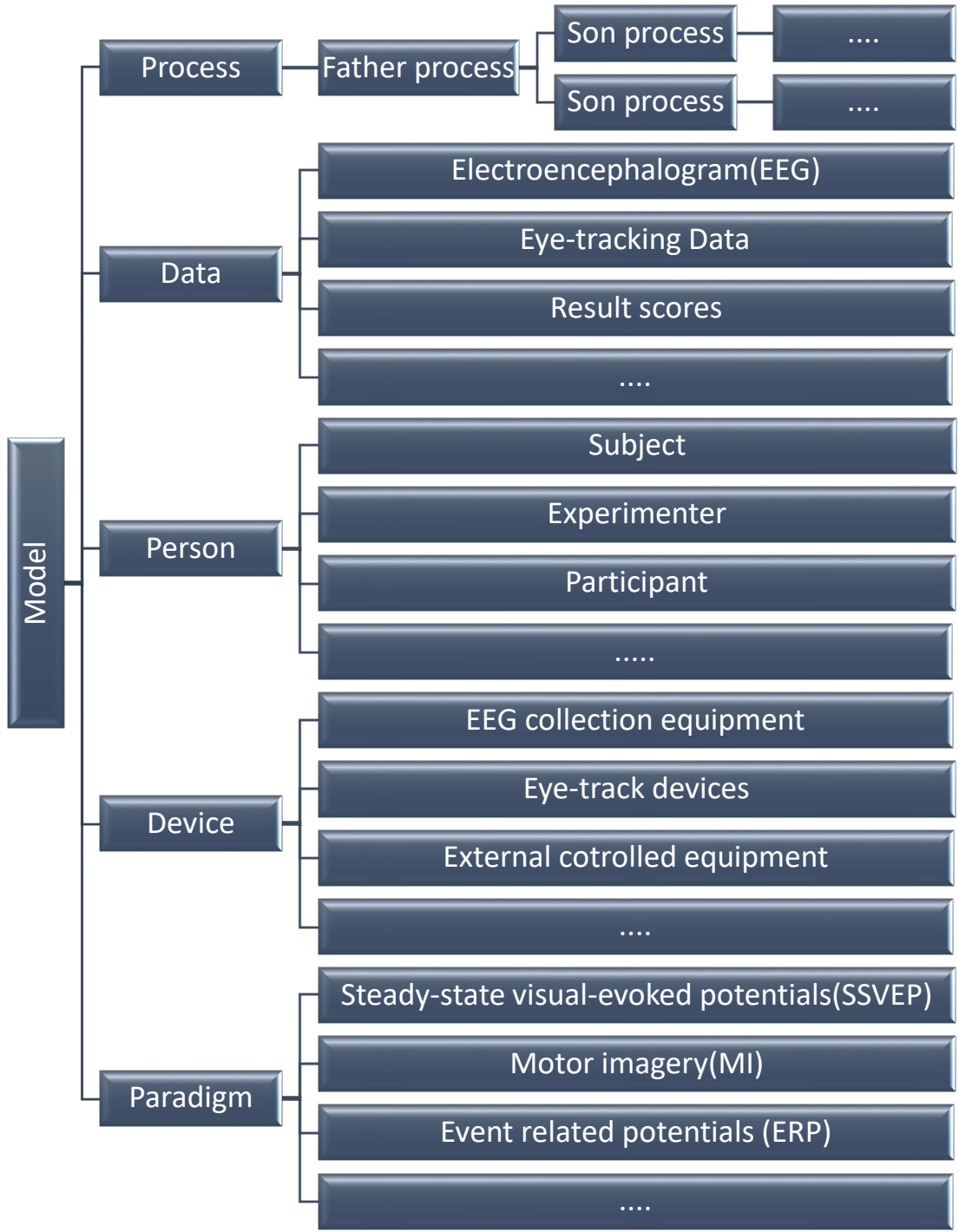

Fig. 1 The content of the data model.

To describe the relationship between data entities under different topics, Entity-Relationship (E-R) diagrams are used to represent these relationships, as illustrated in Fig. 2. The Process entity encompasses all operations that occur in a process into a specific concept, while the Data entity is the main focus of data storage and management. As such, the Process entity and Data entity are positioned at the center of the E-R diagram and are associated with all other data entities. The Process entity has a many-to-many relationship with the Data entity, meaning that a single research process can produce multiple data copies, and a single data copy can be used in multiple research processes, thereby expanding the model's usage scenarios. Both Device and Person can generate Data, and this relationship is crucial information in research and applications. The Paradigm topic is widely studied in BCI research [22], and while Paradigm does not directly produce data, it is an essential part of the experimental process, underscoring the indispensability of the Process concept.

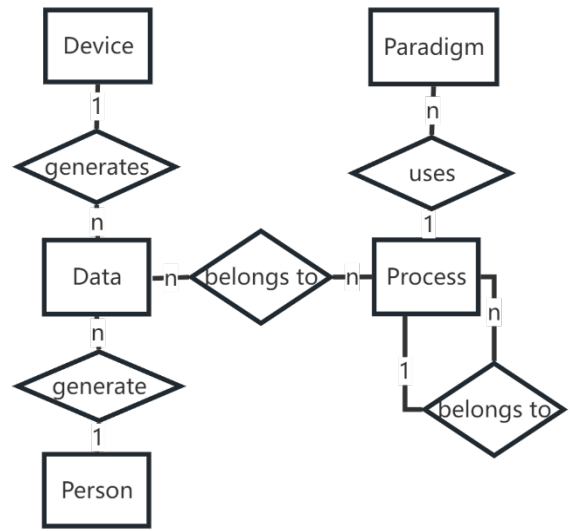

Fig. 2 The relationship of the data model.

## 3.2 Architecture Design

The architecture of NeuroStore is depicted in Fig. 3. It consists of four main components: the data collection layer, the data parsing layer, the data storage layer, and the application layer.

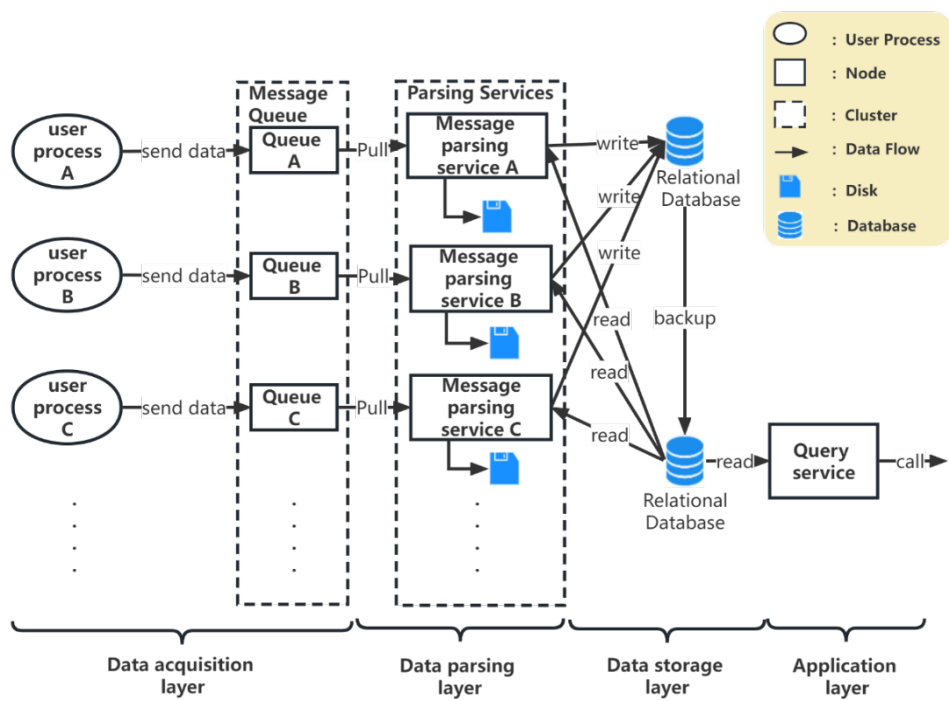

Fig. 3 System architecture diagram.

The data collection layer is responsible for gathering data from endpoints. It provides an API for message sending that allows users to send data from terminals to the system, utilizing automatic connection services and message sending protocols. The data is transmitted over a message queue [23, 24], which is a distributed system that serializes and transmits the data in small packets, ensuring the orderly flow of messages and enabling high concurrency and scalability.

The data parsing layer processes the collected data by transforming it into a standardized format and storing it in a relational database [25]. This layer establishes relationships between data entities, drops unstructured information, and improves the performance and availability of the parsing service by deploying multiple nodes in a service cluster. The relational database is designed based on the common data model.

The data storage layer manages the connections between the upper layer and the database, ensuring the logical correctness of the data relationships in the relational database. It uses a master-slave backup structure [26] for the relational database, where the slave backup not only provides data backup but also serves as a data query service to enhance the query performance. This layer abstracts the underlying database software, facilitating the expansion of the system.

The application layer provides data query and read services to users through a series of query interfaces. This layer decouples the query service from the system, enabling the extension of the query interface in the future.

In conclusion, NeuroStore's architecture is designed to efficiently collect, process, and manage large amounts of data in BCI research. The system provides a flexible and scalable platform for data management, enabling researchers to obtain and analyze data with ease.

### 3.3 API Design

NeuroStore employs an object-oriented design for its interfaces, as depicted in Fig. 4. The design of these interfaces is divided into five sections based on topics, each of which includes storage and query APIs

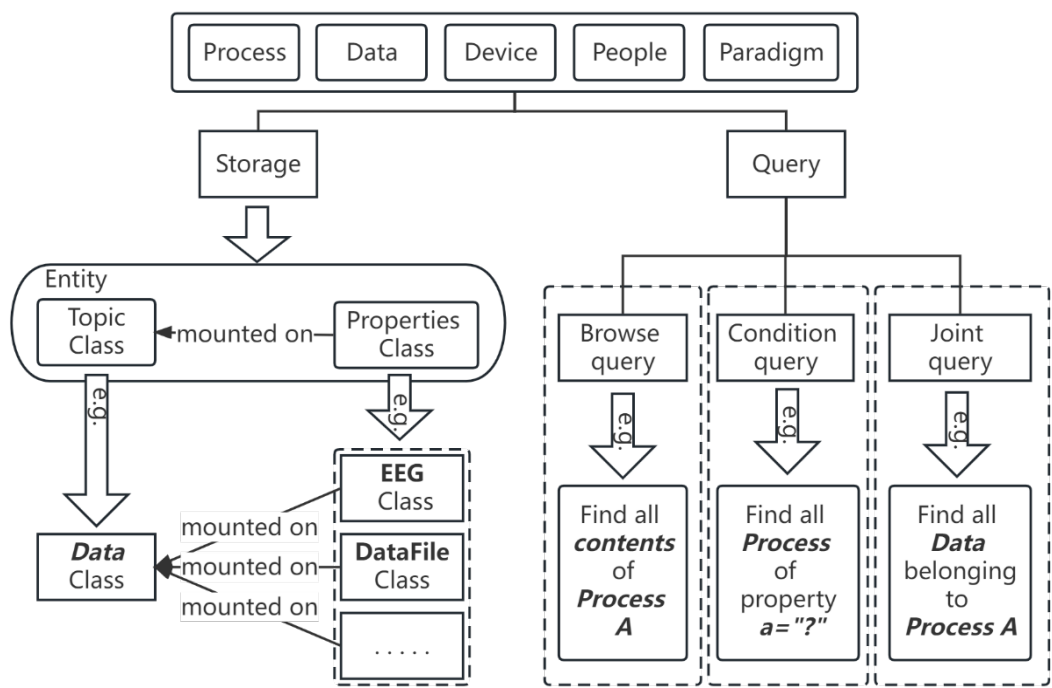

Fig. 4 API schematic.

To utilize the store API, the data to be stored needs to be encapsulated and instantiated as data entity objects according to the relevant data model class specification. The data model class consists of a topic class and attribute class, where each topic has a single topic class (e.g., the Data Class under the Data topic), and multiple attribute classes to describe different data attributes (e.g., EEG Class, DataFile Class under the Data topic). The Topic Class houses mounting functions that can be mounted with one or more attribute classes. This mounting approach enables the combination of the Topic Class and various Property Classes into distinct data entity models, thus increasing the versatility of the interface.

The query API is divided into three main types: browsing query, conditional query, and joint query. The browsing query can be used to query all data under a topic or the details of a specific data, the conditional query can be used to retrieve the data of a topic that meets specific conditions, and the joint query can be used to query data related to other topic data. The three query APIs can be combined and nested to achieve multiple query methods and address different levels of query requirements.

The API design of NeuroStore adopts a classification and combination approach, which reduces the complexity of use and expands the interface function. At the same time, the data storage and query process is standardized through strict interface definition.

To facilitate the use of NeuroStore API, this study provides a concise operational flow, as shown in Fig. 5. To store data, the user first needs to instantiate a data sender to continuously send streaming data. The data to be stored is then encapsulated into the corresponding data entity object based on the relevant data model class specification. Finally, the data entity object is sent to the NeuroStore platform via the data sender. Similarly, in the query process, the user first needs to instantiate a Requester and then select the specific query function based on their needs and combine it into the desired query. Finally, the function return value is passed to the Requester for querying.

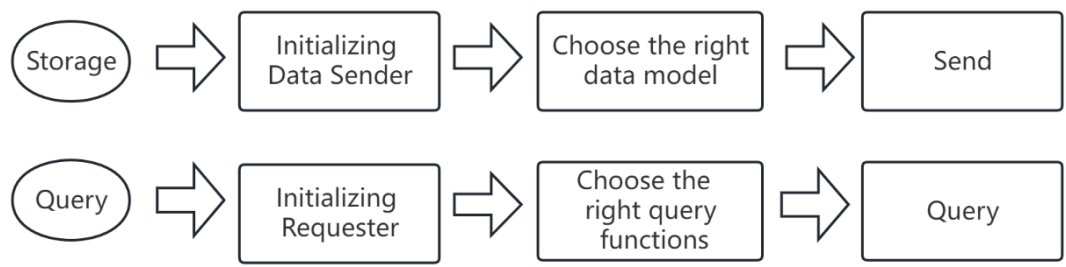

Fig. 5 API usage flow

## 4. Performance Testing
### 4.1 Query Performance Testing
#### 4.1.1 Testing Objectives and Evaluation Metrics

The performance and resource utilization of NeuroStore is rigorously evaluated through testing. The objective of the tests is to determine the system's ability to handle high query loads [27] and to identify any potential performance bottlenecks. The results of the tests provide valuable insight for system tuning and optimization.

Three performance metrics are used to evaluate the system: mean response time, standard deviation of response time, and the p99 [28] response time. The mean and standard deviation of response time provide an indication of the average performance of the system, while the p99 response time represents the response time for 99% of queries, which is a commonly used metric in databases [29, 30] and communication transmissions.

To assess the resource utilization of the system, two key hardware resources are monitored: CPU utilization and memory utilization. These two metrics are typically the most likely to be impacted during high query loads and provide a comprehensive understanding of the system's resource consumption [31].

4.1.2 Test Solution

The evaluation of the performance and resource consumption of NeuroStore was conducted using a test scenario as depicted in Fig. 6. The Jmeter [32] testing tool was utilized, consisting of two components: the Jmeter Manager and Jmeter Server Cluster. The NeuroStore system was divided into two parts: the Relational Database and the Data Persistence Service, with the Relational Database responsible for recording metadata relationships and storing data content, and the Data Persistence Service responsible for processing query requests and converting them into requests to the Relational Database. The dual deployment model was used, with the relational database deployed on one computer and the data persistence service deployed on another computer with the same configuration (16GB of running memory and a 4-core Intel(R) Xeon(R) Platinum 8276M @ 2.20GHz CPU). The Jmeter test cluster and the NeuroStore system under test were located on the same LAN and interconnected via a Gigabit Ethernet network.

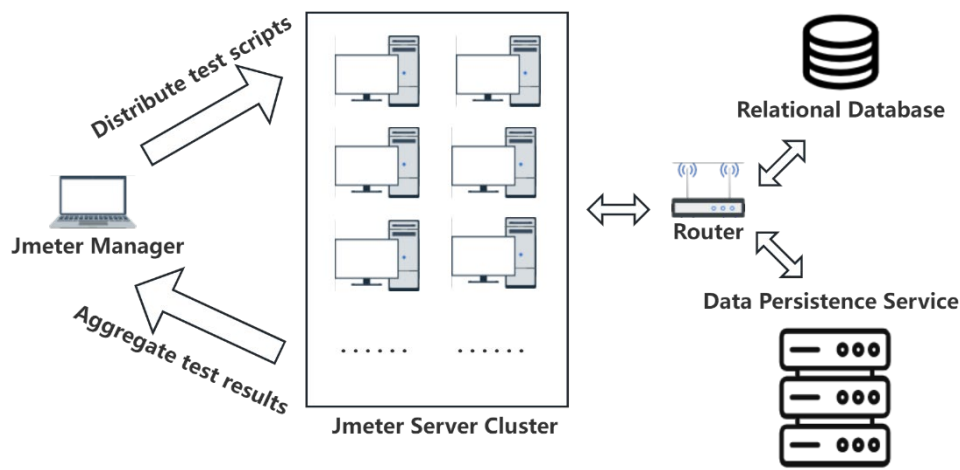

Fig. 6 The query test deployment diagram.

The test script selected the joint query API under the Data topic, which had the longest query time, and each simulated user invoked the interface three times in a loop. The Jmeter Server Cluster simulated high concurrency user access by multi-threading, with all simulated users' threads starting simultaneously in 1 second. The Jmeter Manager automatically collected the response latency of all threads, while the system resource monitoring tool collected the CPU and memory usage of the computer where the Relational Database was located.

The following steps were carried out for the test:
a) 100,000 pieces of randomly generated test data were injected into the system.
b) The total number of simulated users for the test was set to 100.
c) The test script was distributed by the Jmeter Manager to each computer on the Jmeter

Server Cluster, and a start command was sent, causing all scripts to run simultaneously.
d) Upon completion of the test script, the Jmeter Manager was used to collect the query response latency and the system resource monitoring tool was used to collect the CPU and memory usage of the computer where the Relational Database was located.
e) Steps c) and d) were repeated five times, and the average of the five metrics was taken.
f) Steps b) to e) were repeated, increasing the total number of simulated users by 100 each time until 1100 during the repetition.

### 4.1.3 Results Description

As illustrated in Fig. 7, the plot A demonstrates the variation of query response time with an increase in the number of threads representing simulated users. The mean and p99 of response time exhibit a near linear growth with an increase in the number of threads after 200 threads. The standard deviation of the mean value also rises proportionally, indicating that the system stability can result in a response time increase of approximately 400ms for each 100 users. Plot B in Figure 7 depicts the evolution of hardware resource usage as the number of threads grows. The CPU resource usage experiences a rapid increase with the number of threads, which is caused by a high number of concurrent threads accessing the system. However, this does not result in a significant improvement in performance due to the competition for memory resources among multiple threads [31]. The memory usage does not exhibit a significant growth even with the increase in the number of threads, as the database's query engine is configured with a fixed-size buffer pool [33] that is utilized in its entirety. When the number of threads reaches 600, the CPU usage approaches its limit, but the growth rate of the query response time does not change significantly even though the CPU resources are depleted. This is because the test threads are not sending requests continuously, and the number of request congestions is not constantly increasing. In conclusion, the test environment can support concurrent access by up to 600 users if a response latency of no more than 3000ms is required.

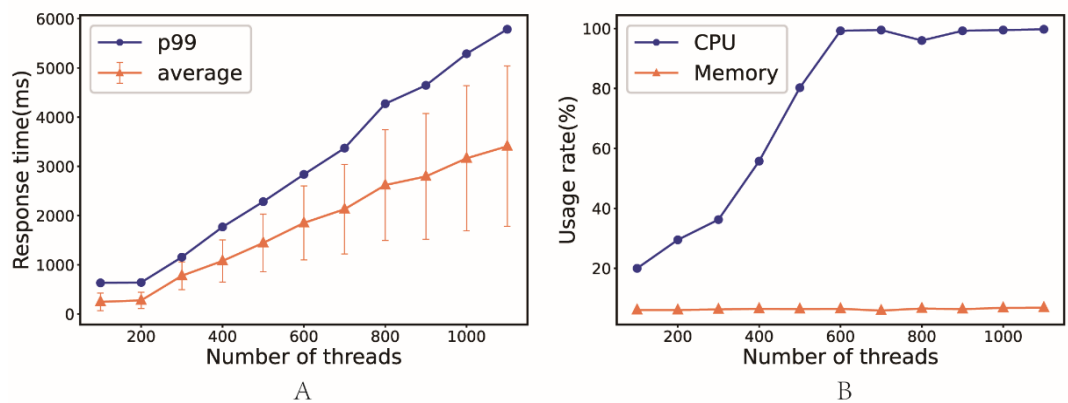

Fig. 7 Query Test Results.

## 4.2 Unstructured Data Storage

### 4.2.1 Testing Objectives and Evaluation Metrics

This test aims to evaluate the data storage capacity of the NeuroStore system, identify any performance bottlenecks, and examine the resource utilization under a specific load. To achieve this, the data storage speed [34] is used as a metric to evaluate the storage capacity of NeuroStore. It is calculated as the ratio of the size of the data files stored on the disk to the test time. As NeuroStore must handle concurrent storage requests and is a disk I/O-intensive software, the CPU utilization and memory utilization [35] continue to be crucial metrics in evaluating the system's operation.

#### 4.2.2 Test Solution

The deployment scheme depicted in Fig. 8 was employed for this test. The NeuroStore system is comprised of a Message Queue and a Data Persistence Service. The Message Queue receives data streams and ensures their order, while the Data Persistence Service records the streaming data, such as EEG data. The Message Queue and the Data Persistence Service were deployed on two separate computers with identical configurations (16GB of memory, a 4-core Intel(R) Xeon(R) Platinum 8276M CPU @ 2.20GHz, and Ubuntu 20.04.4 LTS operating systems, and a Gigabit network card).

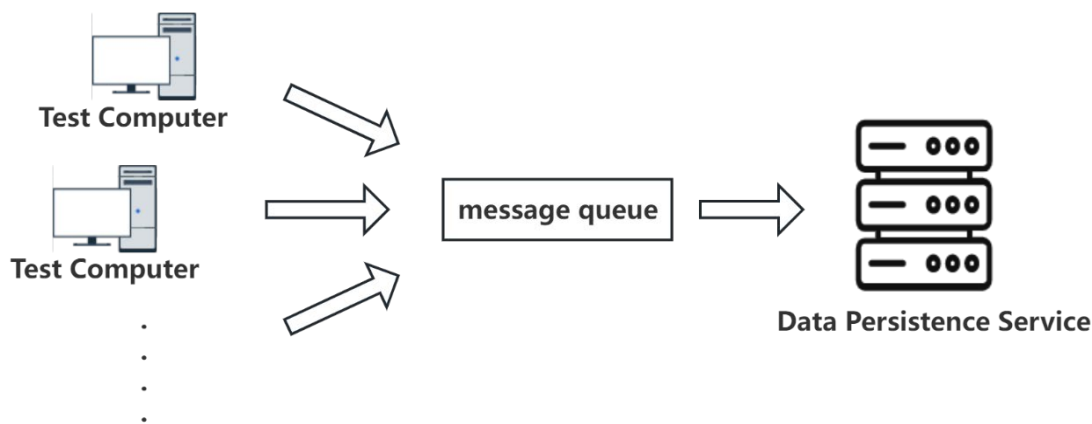

Fig. 8 The storage test deployment diagram.

The test script was designed to simulate the acquisition and storage of EEG data. In EEG data acquisition, a typical acquisition device utilizes 65 channels with a sampling rate of 1000Hz, resulting in 65,000 data points per second, each with a size of 8 bytes.

The test procedure was as follows:
a) The number of simulated EEG data acquisition devices was set to 20, and all test scripts on all Test Computers were set to send data simultaneously.
b) After 10 minutes, all Test Computers simultaneously stopped sending data. The CPU and memory usage of the Data Persistence Service during the test was collected, as well as the size of the stored data on the disk.
c) Steps a) and b) were repeated three times and the average of the three metrics was taken.
d) Steps a) to c) were repeated, increasing the total number of simulated EEG data collection devices in a) by 20 each time until 180 for each repetition.

#### 4.2.3 Results Description

The results of the experiment are presented in Table. 1. As the number of simulated EEG data collection devices increases, the data storage speed increases proportionally. When the number of devices reaches 180, it was expected that the data storage speed would reach approximately 90 MB/s. However, the actual results deviate significantly from this value, suggesting that the storage performance of the system has reached a bottleneck. Concurrently, the CPU resources were nearly fully utilized. Given that the computer is equipped with a Gigabit network card, the theoretical network speed limit is 125 MB/s. Hence, it can be concluded that the system performance is limited by the CPU. This highlights the need for optimization of the system message parsing process and message drop disk to enhance system performance.

Table. 1 Results of test for storage

| Number of Threads | Storage Speed (MB/s) | CPU Usage Ratio(%) | Memory Usage Ratio(%) |
|---|---|---|---|
| 20 | 9.6 | 25 | 9.6 |
| 40 | 19.6 | 36.1 | 10.1 |
| 60 | 29.3 | 42.5 | 10.5 |
| 80 | 39.1 | 51.3 | 9.9 |
| 100 | 49.1 | 62.3 | 12.2 |
| 120 | 59.2 | 71.5 | 11.1 |
| 140 | 68.9 | 78.1 | 10.9 |
| 160 | 78.5 | 89.8 | 11.3 |
| 180 | 86.3 | 98 | 11.4 |

## 5. Application
### 5.1 Application Background

NeuroStore data persistence system was utilized in the BCI Controlled Robot Contest as part of the World Robot Contest (BCI Contest). The main objective of the BCI Contest was to advance the practicality of brain decoding algorithms, through comparison of algorithm performance among different participating teams. The contest involved simultaneous EEG data acquisition from three subjects and the parallel processing of 32 participant algorithms in real-time. The contest was conducted over a period of 4 days and comprised of 5 sub-projects. It required the recording of the entire process, including EEG data, results processed by each algorithm, competition flow, and real-time display of performance ranking and details of each participant.

To effectively manage and store the contest data, a suitable data model with adequate concurrency performance was necessary. In order to meet the requirements of the BCI Contest, NeuroStore was deployed in a containerized manner, consisting of four containers: a message queue, a data parsing service, a relational database, and a result display service. These containers were deployed on a virtual machine with 16GB of RAM and an 8-core Intel Xeon Cascade Lake (2.5 GHz) CPU, running on a cloud server. The data to be stored was written to NeuroStore through programs that utilized its storage APIs. Participants were able to access their results through their browser, with the web backend using the query APIs of NeuroStore to complete the query.

### 5.2 Results Description

During the BCI Contest, NeuroStore received and parsed 79,212 messages and inserted 547,116 data into the data table. Fig. 9 illustrates the message parsing of the last event of the competition, which lasted approximately 4 hours and was divided into four blocks. During the process, there were several spikes in parsing speed. Some of these spikes were due to an increase in business activity, while others were caused by network fluctuations. However, the use of message queues allowed the system to effectively manage these spikes. As a result, NeuroStore was able to store 4.4 Mbit/s of EEG data in real-time and generated a total of 98.4 GB of data files in the BDF format.

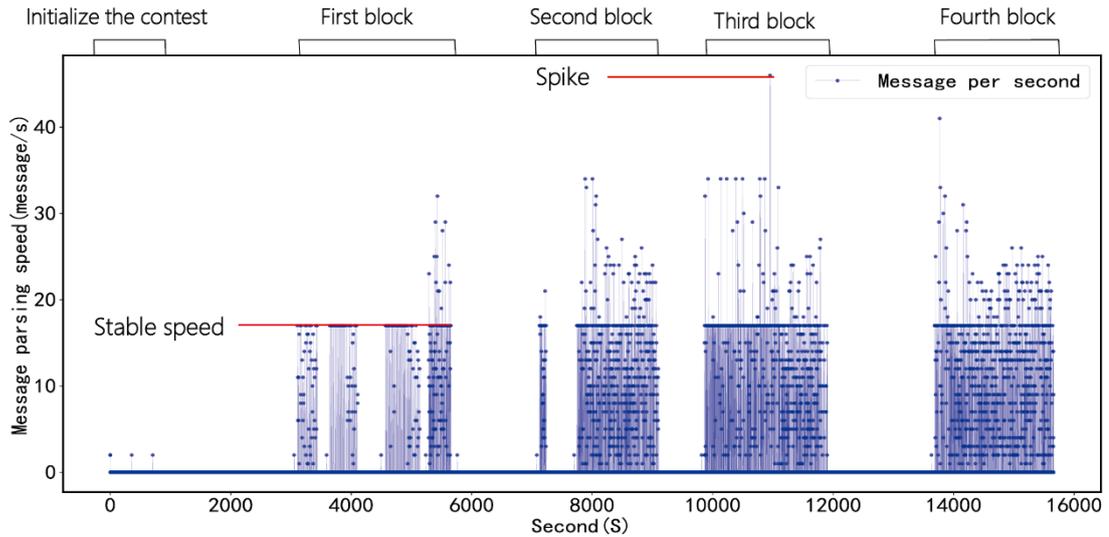

Fig. 9 The process of message parsing.

The results of the contest were displayed in real-time through the interface shown in Fig. 10. This interface displays the performance details of each trial, block, subject, question, and team of the final round of the skills competition. In addition to this interface, there are also performance overview and all-around performance display interfaces, which provide an overview of the subjects' scores.

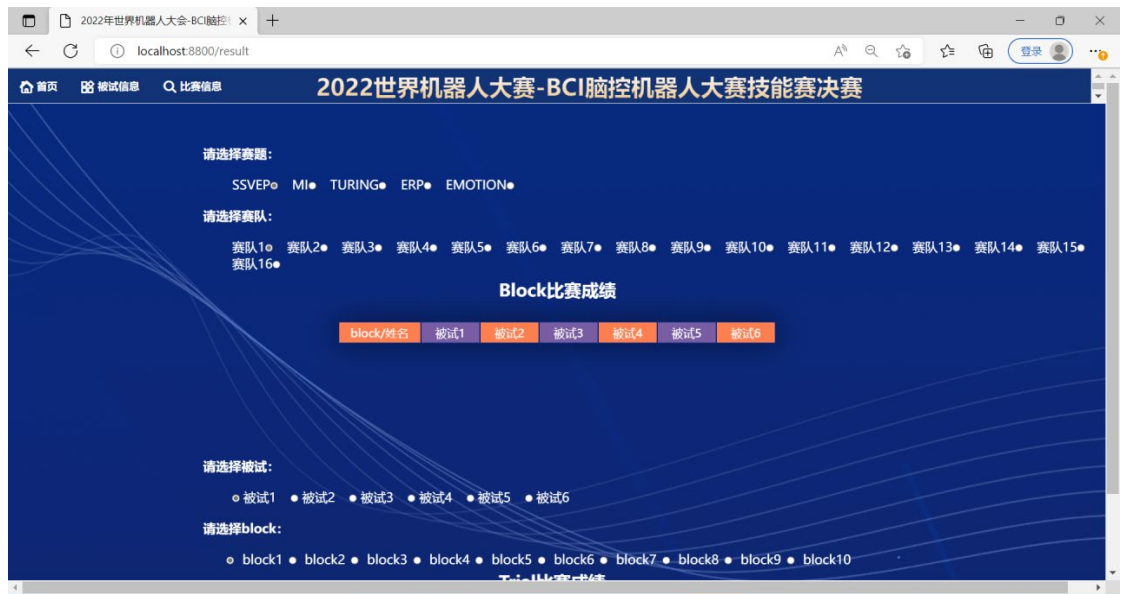

Fig. 10 Results display page.

## 6. Discussion and Future Work

The results of the experiments conducted in 4.1 indicated a relationship between the performance of NeuroStore and its configuration parameters. It was observed that different parameter settings can result in substantial variations in performance, and that each test scenario has its own limitations, thus, cannot provide a comprehensive evaluation of the system. To enhance the practicality of the system, NeuroStore will undertake additional systematic benchmark tests and provide recommendations for parameter configurations in various scenarios.

In terms of versatility, NeuroStore implements existing specifications for data model design. However, as research advances, new data attributes are likely to emerge, requiring updates to the existing data specifications. Hence, the codebase and data model will be continuously maintained and updated to ensure the accuracy of the software and to keep the data specifications up-to-date with the latest application scenarios.

With regards to user-friendliness, this study aims to develop a web interface to navigate and manage the data stored in the database in a more user-friendly manner. Furthermore, interfaces for other programming languages, such as Matlab and Java, will be successively developed to facilitate cross-language usage and to expedite the transformation of research results into practical applications.

## 7. Conclusion

This study presents NeuroStore as a solution to the problem of data storage and management in the BCI domain. The data model in NeuroStore is comprised of five themes, including data entities and their associations, and the system's interface has been designed accordingly, offering versatile data storage and flexible query options. The design of NeuroStore provides a standard solution for BCI data management, aimed at simplifying project data management, promoting data sharing, and increasing the value of data reuse. Furthermore, the system boasts a high-performance architecture. In the test environment, NeuroStore was able to support concurrent access by 600 users with a response latency of no more than 3000ms, and it could also handle real-time storage of data generated by 160 EEG collection devices. As a result, NeuroStore provides a high-performing software implementation that is well-suited for a wide range of scenarios.

# Acknowledgments

This work is granted by the National Natural Science Foundation of China (No 62006024, 62071057), the Aeronautical Science Foundation of China (NO: 2019ZG073001). All the authors thank the subjects who worked for many hours on this project. Thanks to the support provided by the 2022 BCI Controlled Robot Contest in World Robot Contest.